\documentclass[preprint,showpacs,preprintnumbers,amsmath,amssymb]{revtex4}

\usepackage{graphicx}
\usepackage{dcolumn}
\usepackage{bm}

\begin{document}

\title{Phase Synchronization and Polarization Ordering of Globally-Coupled 
Oscillators}

\author{Alessandro Scir\`{e}}
\email{scire@imedea.uib.es}\homepage{http://www.imedea.uib.es/PhysDept}
\affiliation{Instituto Mediterr\'aneo de Estudios Avanzados,
IMEDEA (CSIC-UIB), Campus Universitat Illes Balears, E-07122,
Palma de Mallorca, Spain.}
\author{Pere Colet}
\email{pere@imedea.uib.es}\homepage{http://www.imedea.uib.es/PhysDept}
\affiliation{Instituto Mediterr\'aneo de Estudios Avanzados,
IMEDEA (CSIC-UIB), Campus Universitat Illes Balears, E-07122,
Palma de Mallorca, Spain.}
\author{Maxi San Miguel}
\email{maxi@imedea.uib.es}\homepage{http://www.imedea.uib.es/PhysDept}
\affiliation{Instituto Mediterr\'aneo de Estudios Avanzados,
IMEDEA (CSIC-UIB), Campus Universitat Illes Balears, E-07122,
Palma de Mallorca, Spain.}

\date{\today}

\begin{abstract}

We introduce a prototype model for globally-coupled oscillators
in which each element is given an oscillation frequency and a preferential 
oscillation direction (polarization), both randomly distributed. We found 
two collective transitions: to phase synchronization and
to polarization ordering. Introducing a global-phase and a polarization
order parameters, we show that the transition to global-phase
synchrony is found when the coupling overcomes a critical value
and that polarization order enhancement can not take place before
global-phase synchrony. We develop a self-consistent theory to
determine both order parameters in good
agreement with numerical results.
\end{abstract}

\maketitle
In recent years a considerable interest has been devoted to the
self-organization properties exhibited by networks of coupled
nonlinear oscillators \cite{synchronization}. The work of Winfree
\cite{winfree} first showed that the study of self-sustained
non-identical oscillators is a suitable framework to achieve
insight on the synchronization processes in biological systems.
Based on Winfree's approach, Kuramoto~\cite{kuramoto} proposed a
treatable model for synchronizing oscillators successfully
exploited in many fields, from heart physiology \cite{heart} to
superconducting junctions~\cite{josephson}. The underlying idea
behind this success is that in many instances the dynamics of the
individual oscillators can effectively by described as a limit cycle
in which only one phase plays a relevant role. Then,
for small disorder and weak coupling the Kuramoto model provides
an excellent description of the synchronization process. 
A limitation of this model is that it does not consider
the possible different direction of oscillation of the coupled oscillators.
In fact the relationship between phase synchronization and a
possible collective ordering of the oscillation direction has not been yet
addressed. This question is of direct relevance in the field of optics:
the cooperative behavior encountered in
laser arrays has been investigated both from
experimental~\cite{laser,kapon} and
theoretical~\cite{laser,winful} points of view including
descriptions in terms of the Kuramoto model~\cite{kozyreff} where
the global coupling arises from light feedback from an external
mirror. However, the vectorial nature of the electric field
imposes a fundamental limitation to the description in terms of
single phase oscillators. This description can only be used when
the polarization degree of freedom is completely fixed by natural
constrains. This is not the case, for example, in arrays of
vertical-cavity surface-emitting lasers (VCSELs)~\cite{maxibook},
where the polarization of the emitted light is not fixed by the
structure~\cite{pola}, and the interplay between polarization and
electric field global-phase requires at least a description in
terms of two phases for each element. Indeed, it is possible to have
states in which the global phases are synchronized despite of
a misaligned polarization configuration. Such states have been observed
experimentally in VCSEL arrays~\cite{Debernardi}. Moreover, polarization 
dynamics play an important role in the synchronization of master-slave VCSEL 
configurations, and polarization encoding has been recently proposed for high 
bit-rate encryption in optical communications~\cite{Scire}.

In this Letter, we develop an extension of the Kuramoto model as a
prototype for the study of the fundamental properties of
coupled oscillators described by vector fields in which at least two
phases play a critical role: One associated with
the natural oscillation frequency as in the Kuramoto model, and
the other with the direction of oscillation (polarization). We
study the synchronization properties of an ensemble of
globally coupled non-identical oscillators and show the existence
of two transitions: phase synchronization and
polarization direction ordering. We develop a self-consistent
theory to determine the thresholds for both transitions and show
that polarization ordering can never take place if the
system is not already synchronized in frequency.


Our analysis is made in the context of a general model, the Vector
Complex Ginzburg-Landau Equation (VCGLE), which has been used for modelling
different physical systems, from two-components Bose condensates~\cite{BE} to
non-linear optics~\cite{NLO} including laser emission from wide aperture
resonators such as VCSELs~\cite{vcgle,PRLmaxi}. The VCGLE can be written on
symmetry grounds, but the determination of the parameters in the equation
requires a specific physical model. We consider here parameter ranges of
interest in optics. A set of $N$ globally-coupled space-independent VCGLEs
is given by
\begin{eqnarray} 
\partial_t A_{j}^{\pm} &=& (\mu_j+i\omega_j )A_{j}^{\pm} -(1+i\beta_j)(|A_{j}^{\pm}|^2+\gamma_j|A_{j}^{\mp}|^2)A_{j}^{\pm}
\nonumber \\
&& - (\gamma_{a}+i\gamma_{p})A_{j}^{\mp}e^{\pm i\delta_j}
+\frac{C}{N}\sum_{k=1}^N A_{k}^{\pm}\, , \label{vcgle}
\end{eqnarray}   
where $A_{j}^{+}$ ($A_{j}^{-}$) is the circularly right (left)
polarized component of the $j_{th}$ ($j=1 \dots N$) vector
variable, $\omega_j$ is the natural oscillation frequency,
$\beta_j$ gives a nonlinear frequency shift and $\gamma_j$ (a real number 
for lasers) couples the polarization components.  
The term
$(\gamma_{a}+i\gamma_{p})\exp({\pm i\delta_j})$ represents an
external forcing \cite{PRLmaxi} that linearly couples 
$A_{j}^{+}$ and $A_{j}^{-}$. For example, for a VCSEL, the forcing
arises from device anisotropies (dichroism and birefringence)~\cite{pola} 
that couple the circularly polarized components of the electric
field, introducing two preferential polarization directions.
Another example is a ring laser where any localized change in the
refraction index breaks the invariance along the ring,
introducing the same coupling between the two
counter-propagating modes~\cite{ring} and setting a
preferential phase relationship between them. 
$C$ is the strength of
the global coupling which in laser arrays may be induced by external
reflections (e.g. by placing a reflection at the common focus of
the array~\cite{Arecchi}) or by a common active
medium~\cite{Debernardi}. We introduce $A_{j}^{\pm}=Q_{j}^{\pm}\exp{(i\varphi_{j}^{\pm})}$.
%
We consider $\gamma_j<1$, for which linearly polarized states
($Q_{j}^+=Q_{j}^-$) are stable solutions of the solitary
oscillators~\cite{vcgle}, as it is the case of
VCSELs~\cite{pola}. Close to these solutions, we neglect the
dynamics for each polarization component amplitude
($\dot{Q}_{j}^{\pm}=0$), so that the system (\ref{vcgle}) can be
described in terms of phase equations for each
oscillator: The global phase $\phi_j=(\varphi_{j}^{+} +
\varphi_{j}^{-})/2$, and the rotational phase
$\psi_j=(\varphi_{j}^{+} - \varphi_{j}^{-})/2$, whereas
the latter determines the linear polarization direction. 
We have
\begin{align} 
\dot \psi_j&=\gamma_a \sin(2\psi_j-\delta_j)
+\frac{C}{N}\sum_{k=1}^N\sin
\left(\psi_k-\psi_j\right)\cos\left(\phi_k -
\phi_j\right) \label{kurav1}\,, \\
\dot \phi_j&=\omega_j +
\gamma_p\cos(2\psi_j-\delta_j) 
+\frac{C}{N}\sum_{k=1}^N\sin \left(
\phi_k-\phi_j\right) \cos\left(\psi_k -
\psi_j\right)\,. \label{kurav2}
\end{align}
%
%

In the uncoupled case ($C=0$), the global phases $\phi_j$ rotate
at a constant frequency, whereas the polarization
angles $\psi_j$ reach a steady state, thus modelling a solitary
laser emission. In fact, for $C=0$ we have two orthogonal linearly
polarized solutions for the $j_{th}$ oscillator:
$2\psi_j=\delta_j$, $\phi_j=\phi_{0j}+(\omega_j + \gamma_p)t$, and
$2\psi_j=\delta_j+\pi$, $\phi_j=\phi_{0j}+(\omega_j - \gamma_p)t$,
where $\phi_{0j}$ is a constant. For $\gamma_a<0$ the first
solution is selected, whereas for $\gamma_a>0$
the second is selected. In laser physics, the parameter
$\gamma_a$ models the different linear gain encountered by the two
linearly polarized solutions, thus making linearly stable the solution with
the higher linear gain. In the same context, the parameter
$\gamma_p$ models the cavity birefringence \cite{pola}, which
splits the emission frequency of the two orthogonal linearly
polarized solutions by an amount equal to $2\gamma_p$. In the
following, we take $\gamma_a<0$, so we will refer to
$2\psi_j=\delta_j$ as to the \textit{natural polarization angle}
of each oscillator. Our results, however, do not depend on this
choice, neither do on the sign of $\gamma_p$, which we set
positive. Fixing the polarization degree of freedom
($2\psi_j(t)=\delta_j=\delta_0$ for all $j$) equation
(\ref{kurav2}) reverts to the Kuramoto model: $\dot \phi_j=
\omega_j + \gamma_p + \frac{C}{N}\sum_{k=1}^N\sin(\phi_k-\phi_j)$.
%
%
%

The differences in the natural polarization angles and 
frequencies of the oscillators represent two
sources of disorder in our system. They are statistical
quantities, randomly chosen from two symmetric unimodal
distributions $q(\delta)$ and $p(\omega)$, with zero mean and
standard deviation $\sigma_\omega$ and $\sigma_\delta$,
respectively. Therefore, we introduce two order parameters to
characterize the degree of phase synchronization and
polarization ordering, respectively
\begin{align} 
\eta\exp{(i\chi)} &= \frac{1}{N}\sum_{k=1}^N\exp{(i\psi_k)} \,,\label{op}\\
\rho\exp{(i\theta)} &= \frac{1}{N}\sum_{k=1}^N\exp{(i\phi_k)}\,.\label{op2}
\end{align}
Without coupling $\rho$ averages to zero
while, as $2\psi_j=\delta_j$, $\eta$ accounts for the natural disorder in
the polarization angle. In the continuum limit,
$\eta=\eta_0=\left|\int_{-\pi}^{\pi}\exp(i\delta)q(\delta)d\delta\right|$,
which is non-zero unless $q(\delta)$ is a uniform distribution between
$-\pi$ and $\pi$.

For small coupling the global phases $\phi_j$ are de-synchronized,
which leads the coupling term in the
polarization Eq.(\ref{kurav1})to the vanish. Therefore
each oscillator remains oscillating in its natural polarization
angle. No polarization interaction takes place until the phases
$\phi_j$ start to synchronize. Increasing $C$, two different scenarios
toward polarization ordering and phase synchrony
($\eta$=$\rho$=$1$) are found depending on the relative strength of the
polarization $\sigma_\delta$ and phase disorder $\sigma_\omega$.

\begin{figure}[tb]
\includegraphics*[width=1.0\columnwidth]{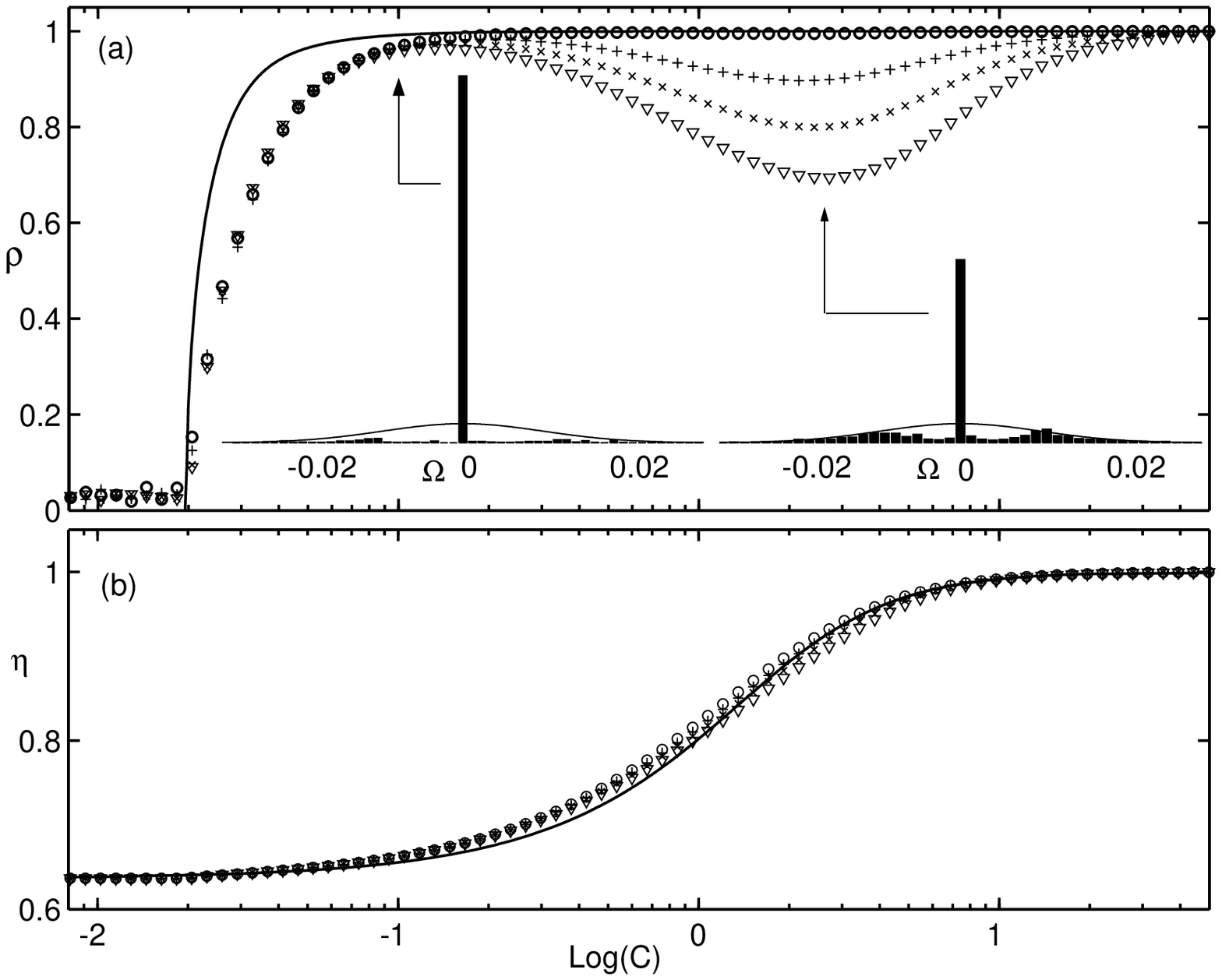}
\caption{Amplitude of order parameters $\rho$ and $\eta$
as function of the coupling $C$. The solid line
corresponds to the theoretical predictions, whereas the
symbols to numerical integration of (\ref{kurav1}) and (\ref{kurav2}) for 
typical VCSEL birefringence values
($\circ$: $\gamma_p=0$, $+$:
$\gamma_p=1$, $\times$: $\gamma_p=2.5$, $\triangle$:
$\gamma_p=5$). We have considered
$\gamma_a=-0.5$, $N=10^3$, a Gaussian distribution $p(\omega$) for the natural
frequencies with $\sigma_\omega=10^{-2}$, and a
uniform distribution $q(\delta)=\frac{1}{2\Delta}$ for 
$-\Delta \leq \delta \leq \Delta$, with $\Delta=\pi/2$
($\sigma_\delta=\frac{\Delta}{\sqrt{3}}=0.9068$) for the natural
polarization angles. The insets show the time-averaged dressed frequencies 
distribution for $\gamma_p=5$, $C=0.1$ (left) and $C=2.5$ (right). The natural 
frequency distribution $p(\omega$) is shown for reference (solid line). 
\label{fig:f1} }
\end{figure}

For $\sigma_\omega\ll\sigma_\delta$, the transitions to phase and
polarization synchrony are well separated. The phases $\phi_j$
synchronize first. The transition to phase synchrony can be
analyzed by taking $2\psi_j=\delta_j$ (frozen polarizations), so
that the set (\ref{kurav1})-(\ref{kurav2}) can be approximated by
\begin{equation} 
\dot \phi_j= \omega_j + \gamma_p +\frac{C}{N}\sum_{k=1}^N\sin \left(
\phi_k-\phi_j\right)\cos\left(\frac{\delta_k -
\delta_j}{2}\right). \label{red0}
\end{equation}
Averaging the polarization angles, reduces Eq.~(\ref{red0}) to a
Kuramoto-like model with an effective coupling $\tilde{C}$
\begin{eqnarray} 
\dot \phi_j&=& \omega_j + \gamma_p +\frac{\tilde{C}}{N}\sum_{k=1}^N \sin
\left(\phi_k-\phi_j\right)\,, \label{red1}
\end{eqnarray}
%
where
$\tilde{C}=C\int\cos\left[(\delta-\delta')/2\right]q(\delta)q(\delta')d\delta=
C\eta_0^2$. The polarization disorder makes the
phase coupling less effective but not vanishing.
Following the standard treatment of the Kuramoto model~\cite{kuramoto}, the
self-consistent equation for the order parameter amplitude $\rho$ reads
\begin{eqnarray}
\rho=\tilde{C}\rho\int_{-\pi/2}^{\pi/2}\cos^2(\phi)q(\tilde{C}\rho\sin(\phi))
d\phi\,.
\label{Kura1}
\end{eqnarray}
Therefore, the critical coupling $C_t$ for the onset of collective
phase synchronization reads
\begin{eqnarray}
C_t=\frac{2}{\pi p(0)\int\cos\left[(\delta-\delta')/2\right]q(\delta)q(\delta')
d\delta}\,.
\label{critical}
\end{eqnarray}
Fig.~\ref{fig:f1} shows the good agreement between the transition
to phase synchronization obtained from numerical integration of
(\ref{kurav1})-(\ref{kurav2}) and the solution of the
self-consistent Eq.~(\ref{Kura1}). Notice the
excellent agreement obtained for the onset of synchronization
given by (\ref{critical}), $C_t=0.01968$.
The distribution of averaged dressed frequencies $\Omega=
\left< \dot{\phi}\right>$ (left inset of Fig.~\ref{fig:f1})
shows a highly dominant peak which comes from the synchronized oscillators
($0.97 N$ in this case). 
Notice also that for $C<C_t$ the polarization order parameter
takes a constant value $\eta_0$ which corresponds to the initial
polarization disorder, in agreement with the assumptions leading
to (\ref{red1}).  


%
Increasing further the coupling strength, the oscillators
leave the respective natural polarization angles and start to
order in polarization. As the phase synchronization has
already been achieved, we are now in the position to develop a
self-consistent theory for the polarization ordering as
follows: Assuming perfect phase synchronization ($\phi_k=\phi_j$),
Eq.~(\ref{kurav1}) becomes
\begin{equation}
\dot{\psi_j}= \gamma_a\sin(2\psi_j-\delta_j) +\frac{C}{N}\sum_{k=1}^N\sin
\left(\psi_k-\psi_j\right)\,. \label{red2a}
\end{equation}  
Since the individual polarization is not a self-oscillating dynamics,
Eq.(\ref{red2a}) is not a Kuramoto-like model. However, from~(\ref{op}) we
have
$\eta\sin(\chi-\psi_j)=\frac{1}{N}\sum_{k=1}^N \sin( \psi_k-\psi_j)$, which
introduced in Eq.~(\ref{red2a}) yields
\begin{equation}
\dot{\psi}_j=\gamma_a\sin(2\psi_j-\delta_j) -C\eta\sin(\psi_j-\chi)\,.
\label{red2}
\end{equation}
The stationary solution $\bar{\psi}_j(\delta,\eta,\chi)$, given implicitly by
%
\begin{eqnarray}
\gamma_a\sin(2\bar{\psi}_j-\delta_j) -C\eta\sin(\bar{\psi}_j-\chi)=0\,, \label{red2ss}
\end{eqnarray}
can be introduced in Eq.~(\ref{op}) to self-consistently
find $\eta$ and $\chi$. In the continuum limit we have
\begin{equation}
\eta\exp(i\chi)=\int\exp(i\bar{\psi}(\delta,\eta,\chi))q(\delta)d\delta\,.
\label{selfcon}
\end{equation}
Altogether, Eqs.~(\ref{red2ss}) and (\ref{selfcon}) allow for the
calculation of the polarization order parameter, for example through a 
Newton-Raphson method, so that the polarization ordering can be fully described. The imaginary part of integral
(\ref{selfcon}) was found to vanish ($\chi=0$), if $q(\delta)$ is even.
Fig.~\ref{fig:f1} shows the agreement between the evaluation of $\eta$ using
the definition (\ref{op}) with the results of the numerical integration of the
full set Eqs.~(\ref{kurav1})-(\ref{kurav2}) and using the self-consistent
approximation given by Eqs.~(\ref{red2ss})-(\ref{selfcon}). We obtain good results
even for small coupling where global phases are desynchronized (in that regime
the contribution of the coupling term in the polarization equation
is negligible).
%
The ordering of the polarization directions induces a loss of coherence  where
the phases partially de-synchronize, lowering $\rho$. The reason is that, as the
polarization order is increased, the polarization angles depart from the natural
angle, and therefore the term $\gamma_p\cos(2\psi_j-\delta_j)$ in
Eq.~(\ref{kurav2}) plays the role of an added disorder to the natural
frequencies $\omega_j$.  Increasing $\gamma_p$, this effect is linearly
increased, enhancing the coherence loss extent, as shown in Fig.~\ref{fig:f1}.
For $\gamma_p=5$, $\rho$ is reduced down to 0.65.
The averaged dressed frequency distribution (right inset of Fig.~\ref{fig:f1}) 
shows that the peak at $\Omega=0$ is lowered in the same proportion and
two lateral lobes associated to drifting oscillators appear, yielding
an overall shape for the distribution similar to that of partially
synchronized Kuramoto oscillators~\cite{kuramoto}. 
From a practical point of view, the coherence lowering would have a direct 
impact the output intensity in VCSEL arrays. A reduction of $\rho$ down to
0.65 leads to coherent output intensity of only $40$\% with
respect to the fully synchronized case. 
Finally, for large coupling, complete phase synchronization
and polarization ordering are achieved.

\begin{figure}[tb]
\includegraphics*[width=1.0\columnwidth]{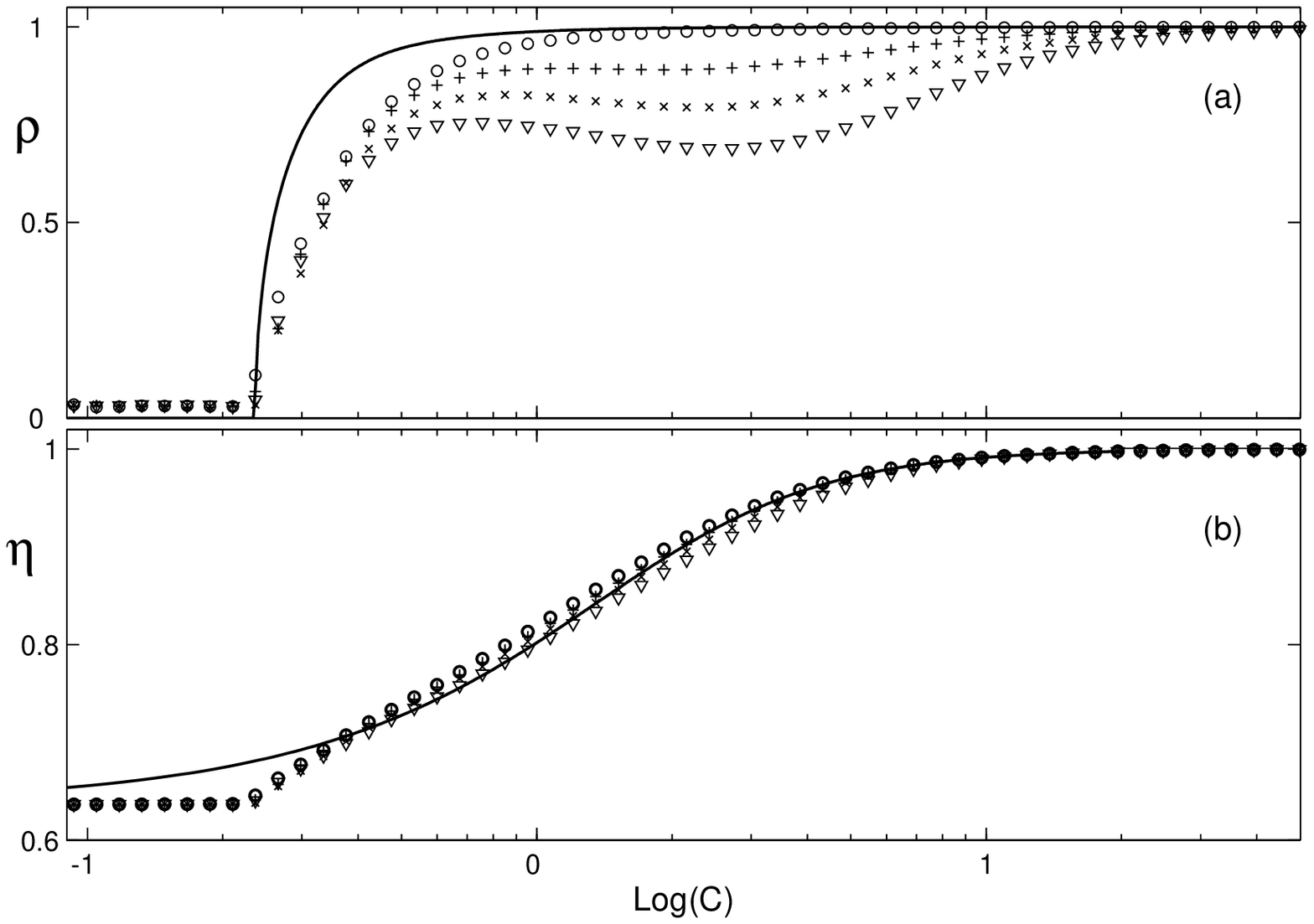}
\caption{\label{fig:f2} Amplitude of order
parameters $\rho$ and $\eta$ as in Fig.~\ref{fig:f1}, but
with larger natural frequency disorder $\sigma_\omega=0.12$, (now $C_t=0.2362$).}
\end{figure}

%
%
Numerical simulations for different values of $\sigma_\delta$ showed that
decreasing the disorder in the natural polarization angles, the
polarization transition to synchronization shifts to lower values
of the coupling. However, a polarization order enhancement is not
possible before the phases start to synchronize, so for
$\sigma_\omega\approx\sigma_\delta$ or
$\sigma_\omega>\sigma_\delta$ the two transitions take place
simultaneously. Nevertheless, the polarizations are still
effectively uncoupled until the phases start to synchronize, so
the self consistent equation (\ref{Kura1}) still holds as well as
the prediction (\ref{critical}) for the phase synchronization
onset $C_t$, which now also signals the onset of the polarization
ordering as shown in Fig.~\ref{fig:f2}. The self-consistent
equation for the polarization order parameter still gives a good
description of the polarization order enhancement.

In conclusion, we have introduced a theoretical framework to study
the synchronization properties of a system of globally coupled
oscillators extending the results for limit cycle oscillators to
include the consideration of oscillation direction (polarization). Two sources of
disorder are included: Randomly distributed natural frequencies
and natural oscillation directions. Increasing the coupling no
polarization order enhancement is possible until the phases start
to synchronize, because the phase disorder destroys the
interaction among the polarization variables. This is in agreement
with experimental results observed in VCSEL arrays
\cite{Debernardi}. Typically, the frequencies synchronize first,
and polarization synchrony takes place at a higher coupling level,
through a partial de-synchronization of the phases (coherence
lowering). We have developed self-consistent approximations which
provide a very good estimation of the  synchronization properties
of system. Increasing the disorder in the natural frequencies or
decreasing the disorder in the natural polarization angle the two
transitions merge in a unique process to full synchrony, and we
provided the critical coupling for its onset.


This work has been funded by the European Commission through VISTA
HPRN-CT-2000-00034, the Spanish
MCyT under project BFM2000-1108, MCyT and Feder SINFIBIO BFM
2001-0341-C02-02.
\pagebreak
%
%

\pagebreak
%
\pagebreak
\pagebreak


\begin{thebibliography}{99}
%
\bibitem{synchronization} S.H.~Strogatz, Nature, {\bf 410}, 268 (2001);
A.~Pikovsky, M.~Rosenblum and J.~Kurths \emph{Synchronization}, Cambridge
University Press, Cambridge UK (2001).
%
\bibitem{winfree} A.T.~Winfree, J.~Theor. Biol. {\bf 16}, 158 (1967);
\emph{The Geometry of Biological time} (Springer-Verlag, New York, 1980).
%
\bibitem{kuramoto} Y.~Kuramoto, in {\it Proceedings of the International
Symposium on
Mathematical Problems in Theoretical Physics}, ed. by H.~Araki, Lecture Notes in Physics, Vol.~39
(Springer, Berlin, 1975); \emph{Chemical Oscillations,
Waves and Turbulence} (Springer, Berlin, 1984); S.H.~Strogatz, Physica D,
{\bf 143}, 1 (2000).
%
\bibitem{heart} C.~S.~Peskin, \emph{Mathematical Aspects of Heart
Physiology}, Courant Institute of Mathematical Science
Publications, NY, (1975).
%
\bibitem{josephson} K.~Wiesenfeld, P.~Colet and S.H.~Strogatz,
Phys. Rev. Lett. {\bf 76} 404 (1996); Phys. Rev. E, {\bf 57} 1563 (1998)
%
\bibitem{laser} L.~Fabiny, P.~Colet, R.~Roy and D.~Lenstra, Phys.
Rev. A {\bf 47}, 4287 (1993). A.~Hardy, E.~Kapon IEEE J.~Quantum
Electron. {\bf 32}, 966 (1996).
%
\bibitem{kapon} H.~Pier, E.~Kapon, and M.~Moser, Nature {\bf 407},
880 (2000).
%
\bibitem{winful} H.G.~Winful and L.~Rahman, Phys. Rev. Lett. {\bf 65}, 1575
(1990); H.G.~Winful Phys. Rev. A {\bf 46}, 6093 (1992); S.~Riyopoulos Phys.
Rev. A {\bf 66}, 053820 (2002).
%
\bibitem{kozyreff} G.~Kozyreff, A.G.~Vladimirov, and P.~Mandel, Phys.
Rev. Lett. {\bf 85}, 3809, (2000).
%
\bibitem{maxibook} D.~Botez and D.~R.~Scifres, \emph{Diode Laser Arrays},
Cambridge Univ. Press U.K., (1994).
%
\bibitem{pola} M. San Miguel in \emph{Semiconductor Quantum
Optoelectronics: from quantum physics to smart devices}, edited by
A. Miller, M.Ebrahimzadeh and D.M. Finlaynson (Institute of
Physics, Bristol, 1999).
%
\bibitem{vcgle}  M.~San Miguel,
Phys. Rev. Lett. {\bf 75}, 425 (1995); E.~Hern\'andez-Garc\'{\i}a,
M.~Hoyuelos, P.~Colet, and M.~San Miguel, Phys. Rev. Lett. {\bf
85}, 744 (2000).
%
\bibitem{Debernardi} P.~Debernardi, G.P.~Bava, F.~Monti di Sopra,
M.B. Willemsen, {\it IEEE J. Quantum Electron.}, {\bf 39}, 109 (2003).
%
\bibitem{Scire} A.~Scir\`e, J.~Mulet, C.~R.~Mirasso, J.~Danckaert, and M.~San Miguel, Phys. Rev. Lett. {\bf 90}, 113901 (2003).
%
\bibitem{BE} R.~Graham and D.~Walls, Phys. Rev. A {\bf 57}, 484 (1998).
%
\bibitem{NLO} M.~van Hecke, C.~Storm, and W.~van Saarlos, Physica {\bf 134D},
1 (1999).
%
\bibitem{PRLmaxi} A.~Amengual, D.~Walgraef, M.~San Miguel, and
E.~Hern\'andez-Garc\'{\i}a, Phys. Rev. Lett. {\bf 76}, 1956 (1996).
%
\bibitem{ring}  M.~Sorel, P.J.R.~Laybourn, A.~Scir\`{e}, S.~Balle,
R.~Miglierina, G.~Giuliani and S.~Donati, Optics Lett. {\bf 27}, 1992,
(2002); R.~J.~C.~Spreeuw, M.~W.~Beijersbergen, and J.~P.~Woerdman,
Phys. Rev. A {\bf 45}, 1213 (1992).
%
\bibitem{Arecchi} S.Y.~Kourtchatov, V.V.~Likhanskii, A.P.~Napartovich,
F.T.~Arecchi, and A.~Lapucci, Phys. Rev. A {\bf 52}, 4089 (1995).
%
\end{thebibliography}
\end{document}